\documentclass[12pt]{article}
\usepackage{./asp2014}

\aspSuppressVolSlug
\resetcounters

\begin{document}

\title{Spectrophotometric support of spectral observations with the telescope BTA}

\author{V.~Panchuk,$^{1,2}$ Yu.~Verich,$^1$ V.~Klochkova,$^1$ M.~Yushkin,$^1$ and G.~Yakopov$^1$ 
\affil{$^1$Special Astrophysical Observatory, Nizhnij Arkhyz, Karachai-Cherkessia, Russia;  \email{panchuk@ya.ru}}
\affil{$^2$ITMO University, S.-Peterburgh, Russia;}}

\paperauthor{V. Panchuk}{panchuk@ya.ru}{ORCID_Or_Blank}{Special Astrophysical Observatory}{Astrospectroscopy laboratory}{Nizhnij Arkhyz}{Karachai-Cherkessia}{369167}{Russia}
\paperauthor{Yu. Verich}{yu.verich@gmail.com}{ORCID_Or_Blank}{Special Astrophysical Observatory}{Observations support laboratory}{Nizhnij Arkhyz}{Karachai-Cherkessia}{369167}{Russia}
\paperauthor{V. Klochkova}{valenta@sao.ru}{ORCID_Or_Blank}{Special Astrophysical Observatory}{Astrospectroscopy laboratory}{Nizhnij Arkhyz}{Karachai-Cherkessia}{369167}{Russia}
\paperauthor{M. Yushkin}{maks@sao.ru}{ORCID_Or_Blank}{Special Astrophysical Observatory}{Astrospectroscopy laboratory}{Nizhnij Arkhyz}{Karachai-Cherkessia}{369167}{Russia}
\paperauthor{G. Yakopov}{yakopov@sao.ru}{ORCID_Or_Blank}{Special Astrophysical Observatory}{BTA technical support team}{Nizhnij Arkhyz}{Karachai-Cherkessia}{369167}{Russia}

\begin{abstract}
We report on the development of the medium--resolution spectrophotometer to accompany (in real time) spectral 
observations performed with the high resolution spectrographs on the 6-m telescope BTA.
\end{abstract}

\section{Description of the spectrophotometer}

We report on the development of the medium--resolution spectrophotometer to accompany  (in real time)  spectral 
observations performed with the high resolution spectrographs~\citep{NES, MSS}  on the 6--meter telescope BTA of the 
Special Astrophysical Observatory.  The spectrophotometer uses a 0.7--meter optics of the BTA guide telescope 
and have a separate system for object positioning.  The spectrophotometer is managed via the Internet. 

In high resolution (R) spectral observations on the BTA the size of the spectral region simultaneously recorded  
depends mainly on the format of CCD used. For example, the  spectral region simultaneously recorded  with 
the Nasmyth echelle spectrograph (NES) is equal approximately to 1000, 1500, 3000\,\AA{} at the CCD format 
1K$\times$1K, 2K$\times$2K, 2K$\times$4K.  Spectral lines  measured in these regions are used for modeling of 
stellar atmospheres by comparing of theoretical and observational  parameters of lines.  
Adequacy of the model  could be checked through comparison both theoretical and observational spectral energy 
distribution. But for such a procedure a photospheric continuum must be registered in broader spectral region, 
i.e. we need low resolution  spectrophotometric observations. For example, spectrophotometric data for stars in 
Pleiades were obtained within 3200$\div$7900\,\AA{} region with  R\,=\,130~\citep{Pleiades}. 
Auxilliary spectrophotometric observations of nonstationary stars must be performed in the equal time  with 
the high and low resolution spectroscopy. Howerever such a regime it is difficult to organize in practice. 

BTA was equipped with the auxilliary reflector having focus 12\,m and diameter of the mirror 0.7\,m  
(Fig.\,\ref{fig1}), which has worked before middle  80--th as a guide system~\citep{Malarev}. 
We analysed  the suitability of this reflector for  spectrophotometric observations making simultaneous 
with high resolution spectroscopy with a main mirror of the BTA.

\articlefigure[width=0.9\textwidth]{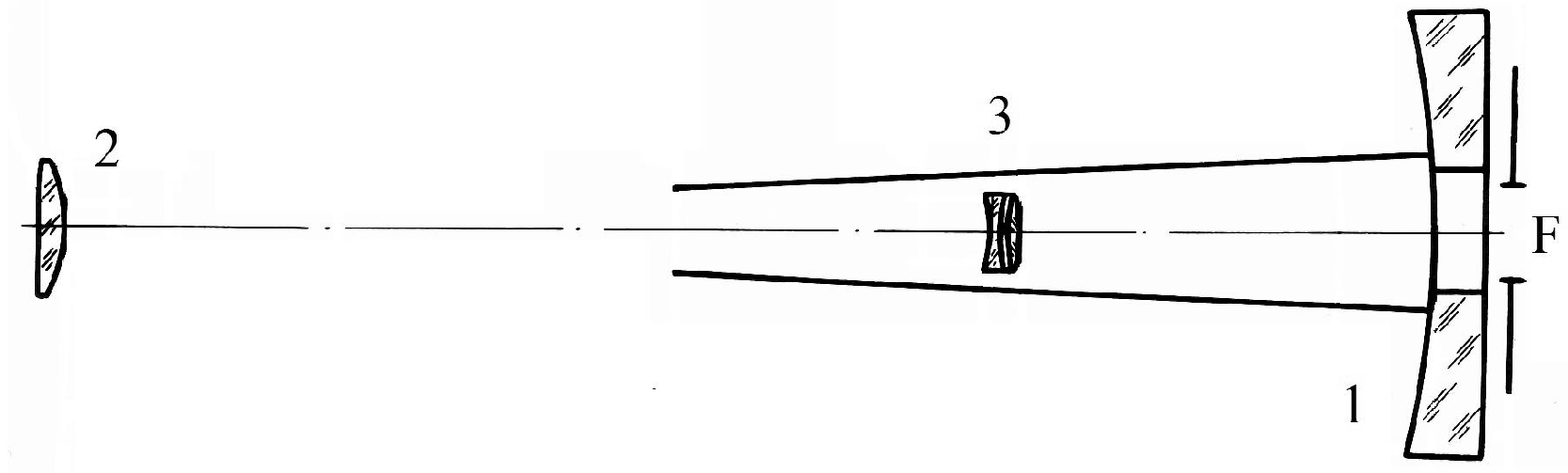}{fig1}{Optics of the guiding telescope of BTA: 1 -- main mirror 
        D$_1$\,=\,0.7\,m,  2 -- secondary mirror D$_2$\,=\,0.21\,m,  3 -- lenses corrector, F -- focal surface.}

The spectrophotometer has to provide the following properties. First, registration time at the 0.7--m telescope 
can not exceed   time registration of high resolution spectrum at the BTA. This condition limits a value of spectral 
resolution of the spectrophotometer. Second, the optics of the 0.7--m telescope has to be achromatic. Therefore we 
removed th two-lenses corrector, after that the effective focal distance of the new layout is equal to 7.8\,m. 
Third, the spectrophotometer has to have an autonomous system to positioning of the star at the input, which has to being 
independent  of the actions  of the observer at the BTA. The layout developed is shown in Fig.\,\ref{fig2}.

\articlefigure[width=0.9\textwidth]{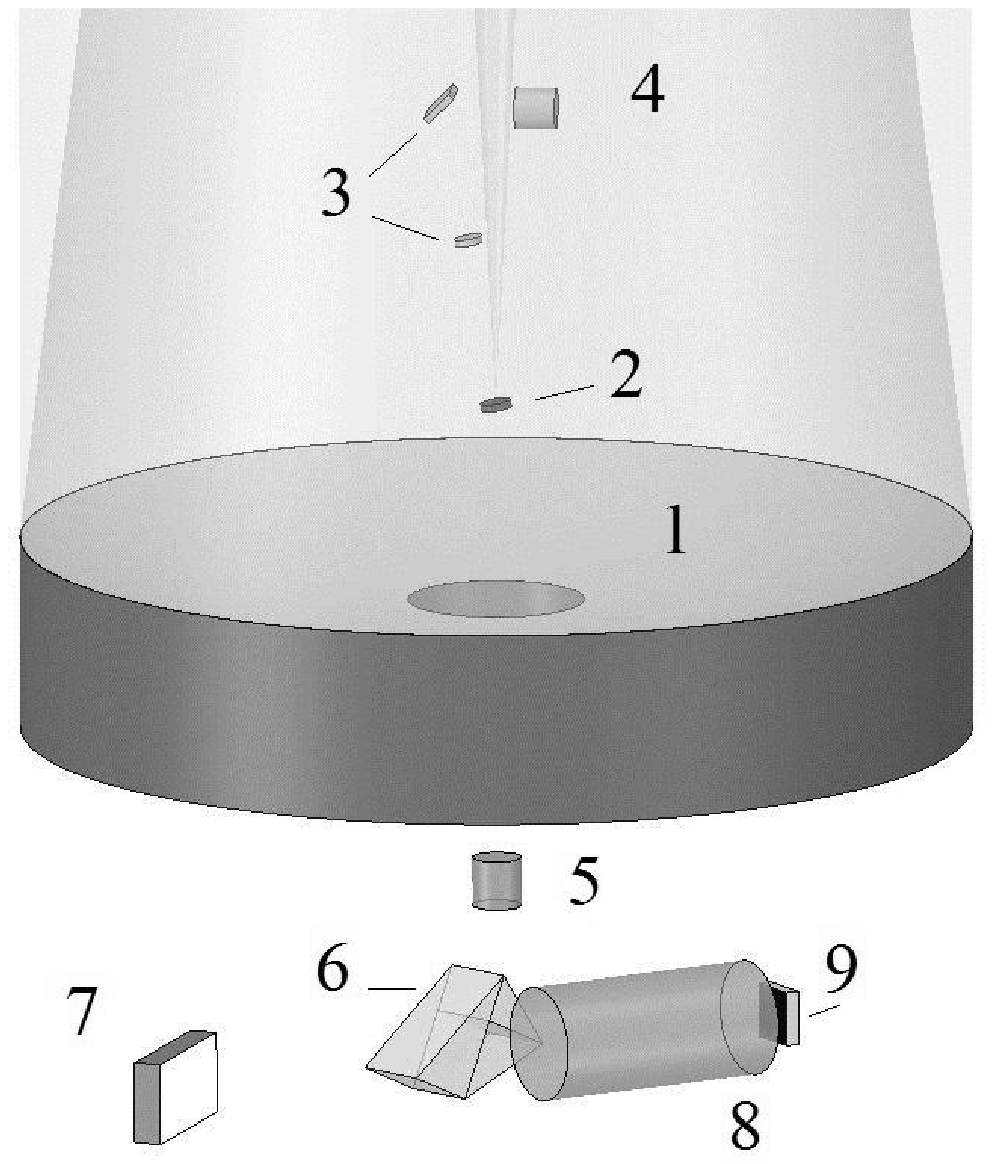}{fig2}{Optical layout of 0.7--meter  reflector in combination with the 
                 spectrophotometer.  
                 1 --  0.7--m mirror,  2 -- dekker at the focus of the reflector,  3 -- optics for guiding of the dekker, 
                 4 -- TV--camera, 5 -- optics of the collimator,  6 -- the cross--disperser  prism, 7 -- echelle,  
                 8 -- lenses camera, 9 -- CCD chip. The secondary mirror is not shown.}

The dekker's size was selected aiming to catch a star image distorted by atmospheric dispersion. The diameter of the 
collimated beam  is equal to 32\,mm. As a cross--disperser the Abbe prism is used which allows to place main elements 
of the spectrophotometer in the plane parallel to back plane of reflector mirror.  We used the echelle  grating 
R2 with grooves density 75\,gr/mm  and the objective  f\,=\,120\,mm.
 
As a local corrector a planeparallel detail ruled by bending in two perpendicular axes is used~\citep{corrector}.  
When a star image wholly get into the dekker, the spectrophotometer works as slitless and spectral resolution is 
determined mainly by accuracy of the star retention on the dekker. Therefore we use a second layout of guiding 
-- inside the spectrophotometer~\citep{guid}. Optical elements of this second layout of guiding are not shown in 
Fig.\,\ref{fig2}.

Results of experiences  of the spectrophotometer and guiding systems will be published separately.

\acknowledgements 
We acknowledge financial support by the Russian Foundation for Basic Research (projects 12--07--00739 and 13--02--00029, 14--02--00291). 


\end{document}